\documentclass[prb,twocolumn,showpacs,amsmath,amssymb]{revtex4}
\usepackage{graphicx}
\usepackage{tabularx}
\usepackage{bm}

\makeatletter
\AtBeginDocument{\@ifpackageloaded{natbib}{\ifNAT@numbers\if@filesw\immediate\write\@auxout{\string\global\string\NAT@numberstrue}\fi\fi}{}}
\makeatother
\begin{document}


\title{Statistics of vortex loops emitted from quantum turbulence driven by an oscillating sphere}


\author{Ai Nakatsuji,$^1$ Makoto Tsubota,$^{1,2}$ and Hideo Yano$^1$}
\affiliation{$^1$Department of Physics, Osaka City University, 3-3-138 Sugimoto, Sumiyoshi-ku, Osaka 558-8585, Japan  \\
$^2$The OCU Advanced Research Institute for Natural Science and Technology (OCARINA), Osaka City University, 3-3-138 Sugimoto, Sumiyoshi-ku, Osaka 558-8585, Japan}

\date{\today}

\begin{abstract}
We perform numerical simulation of quantum turbulence at zero temperature generated by an oscillating sphere. In this simulation, we injected vortices on the sphere to generate turbulence. Although we prepare injected vortex loops of identical length, they are extended by the oscillating sphere to form a tangle through numerous reconnections. The resulting tangle around the sphere is anisotropic and affected by the oscillation. The vortex tangle continues to emit vortex loops, which leave the sphere. The statistics of emitted loops  differ significantly from those of the original injected vortices. First, the sizes of the emitted loops are widely distributed, ranging from smaller to much larger than the size of the initial injected loop. Second, the propagation direction of the emitted loops exhibits  anisotropy: Small loops move away almost isotropically but large ones do so anisotropically along the oscillation direction of the sphere. Thus, the oscillating object stirs the initial injected vortices to reproduce a group of vortices with different statistics. Such physics is compared with the experiments of vibrating objects.
\end{abstract}


\maketitle

\section{\label{sec:intro}INTRODUCTION}
Superfluid helium is a typical quantum fluid that is much influenced by  quantum effects. As a result, quantized vortices appear in superfluid helium. A classical fluid allows vortices to have arbitrary circulation; in contrast, the circulation around a vortex core is quantized in superfluid helium. A tangle of such quantized vortices yields quantum turbulence (QT), which is turbulence of the superfluid component. QT is one of the central issues in low-temperature physics.\cite{book}

To characterize QT, many types of statistical quantities have been studied (e.g., the energy spectrum and the vortex line length). The energy spectrum in fully developed QT obeys the Kolmogorov $-5/3$ law, which is the most important statistical law in classical turbulence.\cite{book} The vortex line length shows how QT develops in a sample volume. These quantities are obtained after performing a coarse-grained average over some macroscopic scale, which does not distinguish any individual vortex. Since QT consists of quantized vortices, there must be statistics that can directly account for each vortex. Such statistics were recently observed in experiments using vibrating wires by the Osaka City University (OCU) group.\cite{Kubo}$^,$\cite{Yano}

Hashimoto {\it et al.} experimentally showed that turbulence starts with remnant vortices attached to wire surfaces in superfluid $^4$He at very low temperature.\cite{remnant1}
They prepared two kinds of vibrating wires: one holding remnant vortices works as a generator of turbulence and another free from them works as a detector.
The turbulence around the generator emits vortex loops, which propagate to the detector.
Recently, a series of experiments was performed at low temperatures by the OCU group.\cite{Kubo}$^,$\cite{Yano} They repeatedly measured the period from the onset of turbulence generation by a generator wire to the vortex detection by a detector wire and found that the non-detection probabilities are well fitted to an exponential distribution, indicating a Poisson process.
In this method, only the first vortex loop reaching the detector was observed.  Using this distribution, they estimated the non-detection time and the mean detection period.
They estimated the size of the first vortices from the non-detection time by assuming they were circular vortex rings.
By measuring the mean detection periods for two different distances between the generators and the detector, the anisotropic emission of the vortices was suggested.

To understand the statistics of each vortex, we need numerical simulations supposing the situation of the vibrating object. There are a few numerical works in which QT made by a vibrating object has been studied.
H$\ddot{\rm a}$nninen {\it et al.} showed how the remnant vortices in superfluid $^4$He become unstable around a sphere under a sinusoidal oscillating flow.\cite{bigsph} The radius of the sphere was 100 $\mu$m and remnant vortices joined the sphere and the vessel wall.
The oscillating flow causes the remnant vortices to excite Kelvin waves and emit vortex loops. The vortex loops form a vortex tangle around the sphere.
Fujiyama {\it et al.} simulated the detector wire in the experiments of the OCU group by using an oscillating sphere.\cite{Fuji2008}$^,$\cite{Fuji2009} They showed how turbulence develops around an oscillating sphere without remnant vortices.
The sphere is hit by vortices injected toward the sphere assuming that the vortices come flying from the generator. The motion of the sphere causes these vortices to be stretched to form a tangle.
These numerical works address course-grained quantities such as vortex line length but do not provide statistics of each vortex.

The main purpose of this paper is to study the statistics of each vortex emitted from turbulence in relation to the OCU experiments.
The OCU group observed only the first vortex loop reaching the detector wire,\cite{Kubo}$^,$\cite{Yano} but we investigate all emitted vortex loops in our simulations.
We focus on the statistics of each vortex, including the length and the direction of the drift velocity.
The direction of the drift velocity may become anisotropic through two mechanisms:  the sphere oscillation and  the configuration of remnant vortices. By using the present method without anisotropy for setting remnant vortices, we can remove the latter mechanism and study the effect of the oscillation.
In  previous works, some specialized forms of remnant vortices were presumed,\cite{bigsph}$^,$\cite{Fuji2008}$^,$\cite{Fuji2009} and the results may depend on these choices.
Here, we add vortices isotropically on the sphere at some intervals; the positions of injected vortices on the oscillating sphere and their normal directions are random.
We also investigate coarse-grained quantities such as vortex line length and the anisotropic parameters.
 We perform the full Biot-Savart vortex filament simulations at zero temperature.

The contents of this paper are as follows. We clarify the configuration and formulation of the model and introduce the equations of the motion in Sec. \ref{sec:eq}. In Secs. \ref{sec:vl} and \ref{sec:ap}, we present the coarse-grained quantities. In Sec. \ref{sec:vl}, we show the vortex line length as a function of time, which illustrates the development of turbulence. In Sec. \ref{sec:ap}, we show the anisotropy of the configuration of vortices around the oscillating sphere. In Sec. \ref{sec:emitted}, we present the statistics of emitted vortex loops, and in Sec. \ref{sec:1.5}, we discuss their dependence on the parameters of an oscillating sphere. Section \ref{sec:conclusion} is devoted to our conclusions.

\section{\label{sec:eq}MODEL AND FORMULATION}
Superfluid $^4$He at zero temperature can be treated mathematically as an ideal incompressible fluid with no viscosity. Quantum fluids have quantized circulation. In $^4$He, the circulation around a quantized vortex is $\kappa=h/m$, where $\kappa$ is called the quantized circulation, $h$ is Plank's constant, and $m$ is the mass of a $^4$He atom. The core size of a quantized vortex is given by the healing length of superfluid $^4$He, which is approximately equal to the atomic size of $^4$He. Hence, we can apply the vortex filament model to this system. According to Helmholtz's theorem, a vortex filament $\bm s(\xi,t)$ moves with the superfluid:
\begin{equation}
\dot{\bm s}=\bm v_s,
\label{eq:vs}
\end{equation}
where $\bm v_s$ is the superfluid velocity at the point $\bm s(\xi,t)$ and $\xi$ is the arc length along the vortex line. In an infinite superfluid system, the superfluid velocity $\bm v_s$ is equal to the velocity $\bm v_{\omega}$ generated by vortex filaments, which is given by the Biot-Savart law,
\begin{equation}
\bm{v}_{\omega}(\bm r)=\frac{\kappa}{4\pi} \int_{L} \frac{(\bm{s}_{l}-\bm{r})\times d\bm{s}_{l}}{\left|\bm{s}_{l}-\bm{r}\right|^{3}},
\label{eq:bs}
\end{equation}
where $\bm s_l$ indicates the position on a vortex, and the integral is performed over all the vortex lines.\cite{sch85}$^,$ \cite{sch88}

In the presence of a boundary, the superfluid velocity satisfies the boundary condition for an inviscid fluid. Then the superfluid velocity $\bm v_{s0}$ in an finite system with a boundary is written as
\begin{equation}
\bm v_{s0}=\bm{v}_{\omega}+\bm v_{b},
\label{eq:sphere}
\end{equation}
where $\bm v_b$ is the additional superflow needed to satisfy the boundary condition. The boundary condition is written as
\begin{equation}
\bm v_{s0}\cdot\bm n=0,
\label{eq:normal}
\end{equation}
where $\bm n$ is the vector normal to the boundary. Although many possible geometries may be considered for  the oscillating objects, in the present study, for simplicity, we use a sphere for which the method for adding image vortices is understood.\cite{image} To satisfy the boundary condition, image vortices that comprise the velocity field $\bm{v}_{b}$ are added inside the spherical boundary.

In addition, in a reference frame moving with the oscillating sphere, the superfluid velocity $\bm v_{s0}$ should satisfy
\begin{equation}
(\bm v_{s0}-\bm v_{p})\cdot\bm n=0,
\label{eq:potential}
\end{equation}
where $\bm v_p$ is the velocity of the oscillating sphere. To satisfy the boundary condition, we add another term $\bm v_u$ to Eq. (\ref{eq:sphere}):
\begin{equation}
\bm v_{u}=\bm {\nabla} \Phi_{u},
\label{eq:grad}
\end{equation}
\begin{equation}
\Phi_{u}(\bm r,t)=-\frac{1}{2}\left(\frac{a}{|\bm r|}\right)^{3}\bm v_{p}(t)\cdot\bm r,
\label{eq:potential_field}
\end{equation}
where $a$ is the radius of the oscillating sphere, $\bm r$ is the position vector whose origin is the center of the sphere, and $\bm v_u$ is the flow induced by the motion of the sphere.\cite{sch74} Thus, $\bm v_{s0}$ is obtained by adding the contribution of vortex filaments, image vortices, and the potential field of Eq. (\ref{eq:grad}):
\begin{equation}
\bm v_{s0}=\bm v_{\omega}+\bm v_b+\bm v_u.
\label{eq:vel_all}
\end{equation}

A vortex filament is represented numerically by a string of points. The numerical spatial resolution, which is the distance between discrete points on a vortex, is 0.05 $\mu$m. Theoretically, when a vortex encounters another vortex or comes close to the oscillating sphere, reconnection can occur. However, our numerical method with vortex filaments cannot represent the reconnection process itself. Hence, we reconnect vortices when a vortex is very close to another or to the oscillating sphere within the numerical spatial resolution.
We remove small vortices whose sizes are comparable to the spatial resolution.\cite{araki} Removing them acts as  dissipation in this system, which is the usual practice followed in a zero-temperature simulation. However, this procedure has no effect on the large-scale phenomena we are interested in.

\begin{figure}[t!]
\begin{center}
\includegraphics[clip, width=2.2cm]{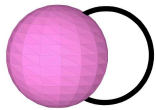}
\end{center}
\caption{(Color online) Injection vortices. The black line shows a vortex whose form is an arc of a circle with a radius of 1.1 $\mu$m and a central angle of $3 \pi /2$ injected onto an oscillating sphere every 0.01 ms. The positions of injected vortices on the oscillating sphere and their normal directions are random.}
\label{fig:step0}
\end{figure}

In this simulation, our current system is in open space. We assume that the surface of the oscillating sphere is smooth. The radius of the oscillating sphere is 1.1 $\mu$m, the frequency of the oscillation is 3 kHz, the amplitude of the oscillation is 5.31 $\mu$m, and the oscillation peak velocity is 100 mm/s. These parameters correspond to the experimental conditions.\cite{Kubo} We inject vortices, whose form is an arc of a circle with a radius of 1.1 $\mu$m and a central angle of $3 \pi /2$, onto the oscillating sphere (see Fig. \ref{fig:step0}).
The parameters of the injected vortex are determined by the parameters of the sphere oscillation.
The motion of the sphere causes an attached vortex to become unstable  when the sphere velocity is greater than the velocity of the vortex of Eqs. (\ref{eq:vs}) and (\ref{eq:bs}). A small vortex attached to the sphere whose velocity is greater than that of the sphere slips over the sphere when the surface of the sphere is smooth. In contrast, a large vortex with a low velocity is detached from the sphere.
The detached large vortex cannot move fast; therefore, the sphere collides with it repeatedly, and the vortex splits into smaller vortices. Thus, we add vortices with a radius of 1.1 $\mu$m that do not really slip on the sphere and can exit the turbulent region because the self-induced velocity of the vortex is comparable to the velocity of the sphere. This length, i.e., 5.18 $\mu$m, is related to the sphere velocity and is denoted by $l_{SV}$. Unless we keep injecting vortices, a turbulent state cannot be maintained. A vortex attached to the oscillating sphere is detached by the sphere motion in about 1 ms; therefore, we inject vortices every 0.01 ms, which is much shorter than 1 ms.
To add these vortices isotropically, the positions and the normal directions of injected vortices on the oscillating sphere are random.

We should study the emission of vortex loops from a statistically steady vortex tangle. However, our vortex tangle is not uniform but localized around the oscillating sphere. Hence, we should develop a criterion for how and where we observe the vortex loops to collect data for statistics. Thus, we suppose a computational spherical region covering roughly the localized vortex tangle.
The radius of the computational spherical region is 30 $\mu$m and its center is set at the center of the sphere oscillation.
First, by monitoring the vortex line length inside the computational spherical region, we investigate how the vortex tangle develops and whether it becomes statistically steady.
Second, we investigate the vortex loops emitted from the vortex tangle in statistically steady state. The emitted vortex loops move away independently when they are far enough away from the tangle around the oscillating sphere. Then, we will detect them at 30 $\mu$m from the center; the statistics observed at 30 $\mu$m is supposed to be the same as that observed by the detector wire that is placed at a distance of the order of 1 mm away from the generator because each vortex is expected to propagate independently out of the computational spherical region.

\begin{figure}[ht!]
\begin{center}
\includegraphics[clip, width=6.0cm]{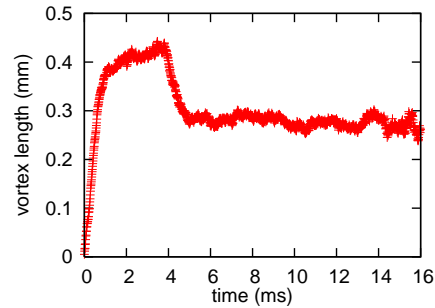}
\end{center}
\caption{(Color online) Vortex line length as a function of time in a computational spherical region.}
\label{fig:vl}
\end{figure}

\begin{figure}[h]
\begin{tabular}{ccc}
\begin{minipage}{0.5\hsize}
\begin{center}
\fbox{\includegraphics[clip, width=3.5cm]{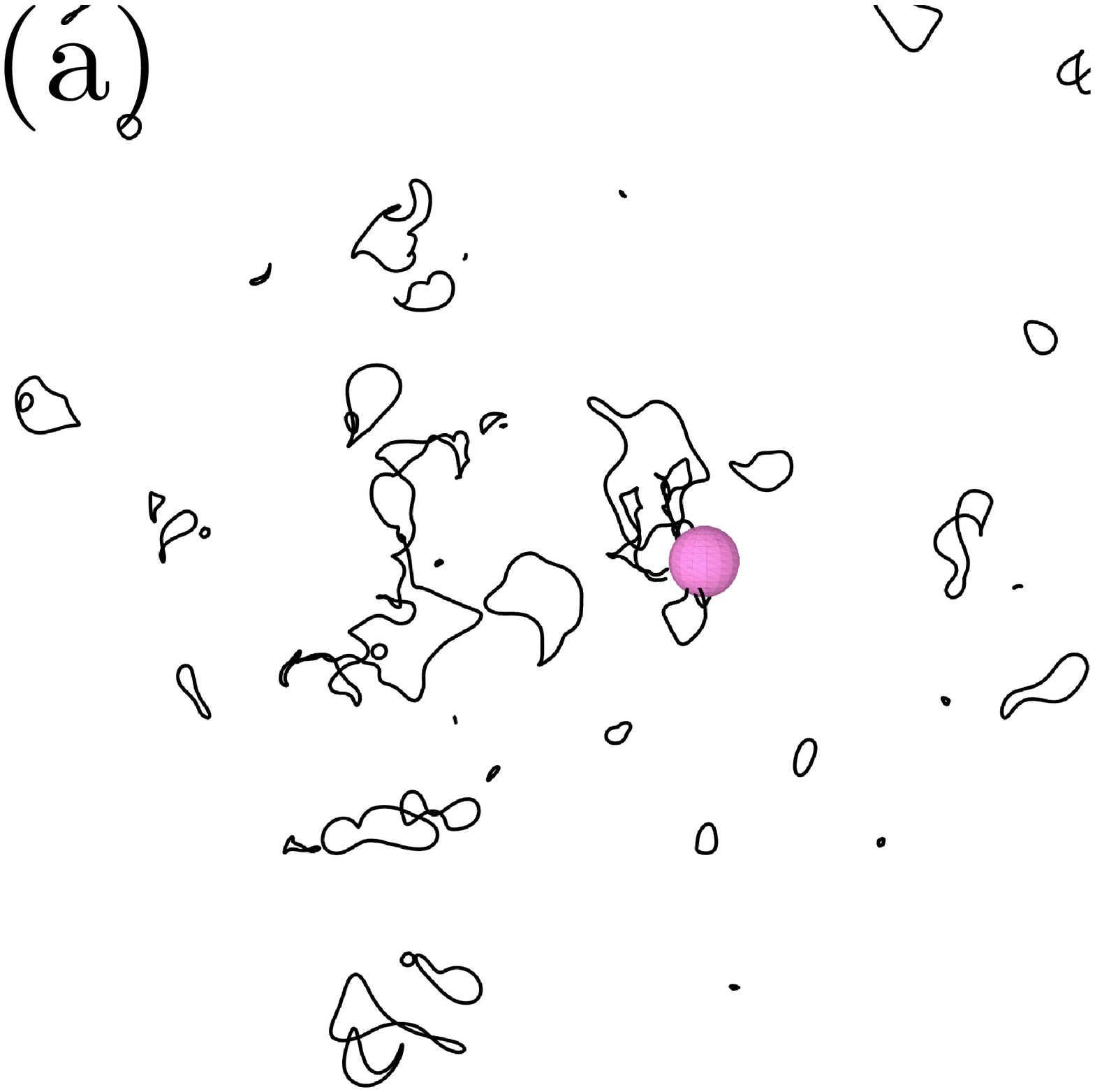}}
\end{center}
\end{minipage}
\begin{minipage}{0.5\hsize}
\begin{center}
\fbox{\includegraphics[clip, width=3.5cm]{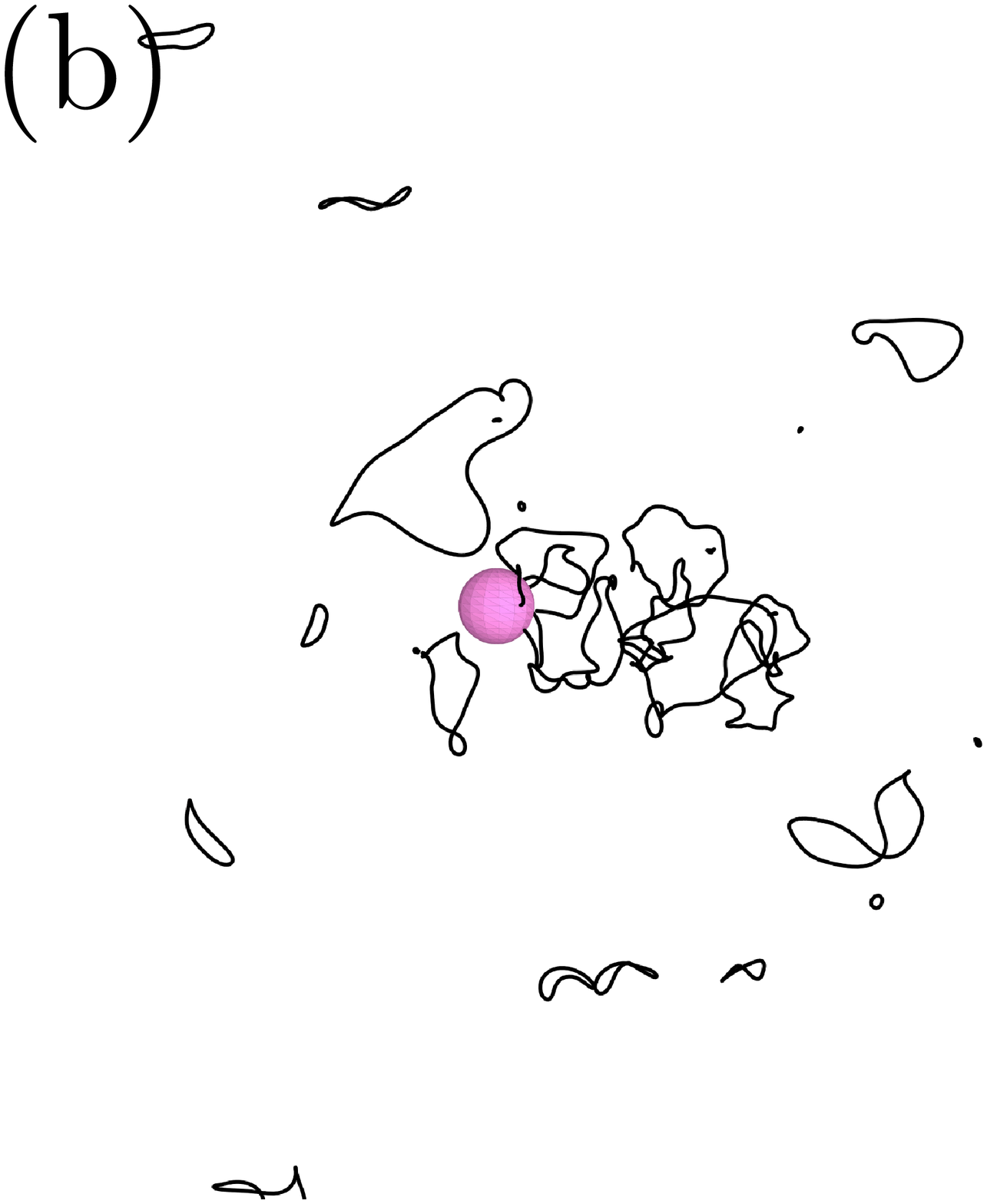}}
\end{center}
\end{minipage}
\end{tabular}
\caption{(Color online) Simulations of the time evolution of a vortex tangle with an oscillating sphere at (a) $t = 2$ ms and (b) $t = 9$ ms.}
\label{fig:snap}
\end{figure}

\section{\label{sec:vl}TIME EVOLUTION OF THE\ GENERATED TURBULENCE}
We now calculate the time development of the vortex line length (VL) inside the computational spherical region (see Fig. \ref{fig:vl}). The VL increases as the vortices are injected and stretched. When vortices exit from the computational spherical region, the VL decreases.

Vortices attached to the oscillating sphere are stretched by the sphere motion. When their edge points on the sphere move toward a stagnation point and approach each other, they reconnect. In this way, the vortex becomes detached from the oscillating sphere and moves away from the sphere. When detached vortices reconnect with each other, vortices of various sizes are generated. What happens to them after that depends on their size. Small vortex loops move off immediately because of its own high drift velocity:\cite{sch88}
\begin{equation}
\bm v_{DF}(t) =\frac{1}{\ell}\int_{c} \dot{\bm s}\  d\xi,
\end{equation}
where $\ell$ is the length of the vortex loop, $\dot{\bm{s}}$ is given by Eq. (\ref{eq:vs}), and the integral is performed over the vortex loop. In contrast, large vortex loops remain around the sphere because of their low drift velocity. The remaining large vortex loops then form a tangle around the sphere and we call this region the turbulent region (see Figs. \ref{fig:snap}(a) and \ref{fig:snap}(b) ).

In Fig. \ref{fig:vl}, for 0  $\leq t <$ 0.95 ms, the VL increases because the injection and the growth of the vortices  dominate over the loss of the vortices. The total length of vortices injected during the period of 0.95 ms is approximately 0.49 mm. The VL at 0.95 ms is approximately equal to the value of the total length.
For 0.95  $\leq t <$ 3.8 ms,  turbulence gradually forms around the sphere, while small vortices escape continuously from the computational spherical region (see Fig. \ref{fig:snap}(a) ).
Then, the increase of the VL tapers off.
For 3.8  $\leq t <$ 5.0 ms, only large vortex loops escape from the computational spherical region, reducing the VL.
After 5.0 ms, in the statistically steady state case, the loss of vortices balances the injection and growth of vortices so that the VL saturates, and the configuration of vortices around the sphere is as depicted in Fig. \ref{fig:snap}(b).

\section{\label{sec:ap}ANISOTROPIC PARAMETERS OF FULLY DEVELOPED VORTEX TURBULENCE}
This system is anisotropic along the direction of the oscillation. Hence, the turbulence in the turbulent region is anisotropic in spite of the isotropic injection of vortices. To show the anisotropy of turbulence, we use the anisotropic parameters.\cite{sch88} The anisotropic parameters are calculated by using the following equations:
\begin{equation}
\bm I_{\parallel}=\frac{1}{L'}\int\left(1-\left(\bm{s}' \cdot \bm{r}_{\parallel}\right)^2\right)d\xi,
\end{equation}
\begin{equation}
\bm I_{\bot}=\frac{1}{L'}\int\left(1-\left(\bm{s}' \cdot \bm{r}_{\bot}\right)^2\right)d\xi,
\end{equation}
where $\bm{r}_{\parallel}$ and $\bm{r}_{\bot}$ stand for unit vectors parallel and perpendicular, respectively, to the direction of the sphere oscillation; $L'$ is the total vortex length in the turbulent region; and the integral is performed over all the vortex lines in the region. The turbulent region is supposed to be a cylindrical region whose radius is 3.3 $\mu$m (three times the radius of the sphere) and whose height is 15.39 $\mu$m (1.5 times the total amplitude of the sphere oscillation). The axis of the cylindrical region is set along the oscillation direction of the sphere. The center of the region is set at the center of the oscillation. If the tangle is isotropic, we have $\bm I_{\parallel}=\bm I_{\bot}=2/3$. However, if vortices are lying in the direction normal to $\bm{r}_{\parallel}$, the direction of the sphere oscillation, we have $\bm I_{\parallel}=1$ and $\bm I_{\bot}=1/2$.

The average values of the anisotropic parameters in the statistically steady state are $\bm I_{\parallel}=0.7106$ and $\bm I_{\bot}=0.6442$.
This indicates that the turbulence in the turbulent region is nearly isotropic but that there are more vortices normal to the oscillation direction than would be the case in purely isotropic turbulence. When we add vortices isotropically at some intervals without an oscillating sphere, isotropic turbulence should be generated. In this condition with the oscillating sphere, the turbulence is somewhat anisotropic. In the previous studies by H$\ddot{\rm a}$nninen {\it et al.}\cite{bigsph} and Fujiyama {\it et al.,}\cite{Fuji2008}$^,$\cite{Fuji2009} the injected vortices were anisotropic. The turbulence around the sphere in these studies could be anisotropic because of the oscillating sphere and the way vortices were injected.

\section{\label{sec:emitted}STATISTICS OF EACH EMITTED VORTEX LOOP}
In this section, we discuss the statistics of each vortex loop emitted from the statistically steady turbulence, focusing on the length of vortex loops, their drift velocity compared with that of a circular vortex ring whose length is identical to that of the emitted vortex loop, and the anisotropy of the propagation direction. The emitted vortices refer to  vortices that can escape from the turbulent region.

\begin{figure}[ht!]
\begin{center}
\includegraphics[clip, width=6cm]{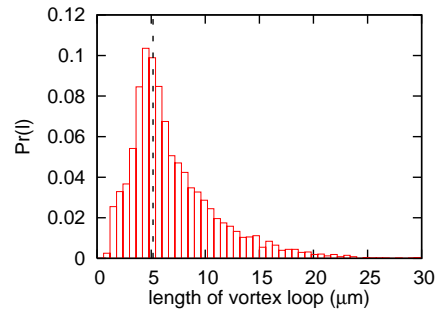}
\end{center}
\caption{(Color online) The PDF of the length of emitted vortex loops. The dashed black line shows the length $l_{SV}$ (5.18 $\mu$m).}
\label{fig:pdf11}
\end{figure}

\subsection{\label{sec:pdf_len}Length of emitted vortex loops}
The lengths of emitted vortex loops are distributed between $\sim$1 and $\sim$30 $\mu$m, even though  the length $l_{SV}$ of the injected vortex determined by the velocity of the oscillating sphere is 5.18 $\mu$m. Such redistribution of the vortex length is generally caused by the two kinds of vortex reconnections, namely the split type and the combination type.\cite{araki} To investigate the size of the vortex loops emitted from the turbulent region, we calculate the length of each emitted vortex loop when it reaches the boundary of the computational spherical region.
Figure \ref{fig:pdf11} shows the probability density function (PDF) of the length of such emitted vortex loops.
The PDF has a peak corresponding to the value of $l_{SV}$.
A vortex whose length is shorter than the original length $l_{SV}$ is generated by the split type of reconnections.
The length of large vortices may depend on the amplitude of the sphere oscillation.
A vortex stretched by the oscillating sphere has a size  similar  to the total amplitude of the oscillating sphere, 10.62 $\mu$m. A vortex longer than $l_{SV}$ is generated by some combination type of reconnections\cite{araki} and stretching processes. Thus, vortex loops of various sizes are created.

\begin{figure}[th]
\begin{center}
\includegraphics[clip, width=6cm]{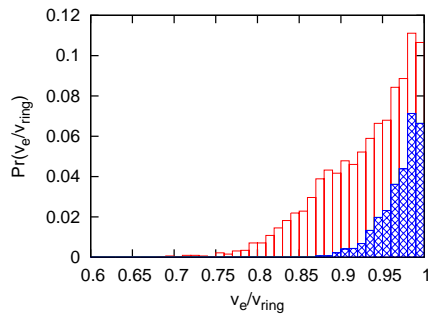}
\end{center}
\caption{(Color online) The PDF of the drift velocity $\bm v_e$ of emitted vortex loops in this simulation normalized by the drift velocity $\bm v_{ring}$ of a circular vortex ring. The blue shaded parts show the rate of small vortices whose length are shorter than $l_{SV}$. The number of data for small vortices is about a third of the whole.}
\label{fig:ve_v}
\end{figure}

\subsection{\label{sec:vel}Drift velocity of emitted vortex loops}
As found in Fig. \ref{fig:snap}, the emitted vortex loops are generally non-circular. Hence the emitted vortex loop has a lower drift velocity  than that of a circular vortex ring whose length is identical to that of the emitted vortex loop. Although the drift velocity of a vortex is time dependent, we obtain the drift velocity averaged for 6 ms after the vortex loop reaches the boundary of the computational spherical region.

Figure \ref{fig:ve_v} shows the PDF of the drift velocity $\bm v_e$ of all emitted vortex loops normalized by that of a circular vortex ring\cite{donelly} with the same vortex length:
\begin{equation}
\bm v_{ring}(l) =\frac{\kappa}{2l} \ln{\frac{4l}{e^{1/2}\pi a_{0}}},
\end{equation}
where $l$ is the length of the vortex ring and $a_{0}$ is the core size of a quantized vortex.

A vortex loop with a Kelvin wave is known to have a lower drift velocity than a circular vortex ring.\cite{PRE}$^,$\cite{kelvin}
A circular vortex has the highest drift velocity among all vortex loops with the same length. Hence, the value of $\bm v_e/\bm v_{ring}$ is less than unity.

The sphere oscillation may excite  Kelvin waves with a wavelength resonant with the frequency of the oscillation. However, in this condition, the wavelength is about 13 $\mu$m. Because this wavelength is generally much greater than the total length of vortices, the excitation of Kelvin waves by the oscillation would be not so important.

In the experiment using vibrating wire performed by the OCU group, the radius of emitted vortex loops was estimated by assuming the vortex loops to be  circular vortex rings. In our simulation, $\bm v_e/\bm v_{ring}$ is distributed  between $\sim$0.68 and $\sim$1 (see Fig. \ref{fig:ve_v}). The PDF of the drift velocity depends on the length of the emitted vortex loops.
The shaded parts in Fig. \ref{fig:ve_v} show the contribution of small vortices whose lengths are less than $l_{SV}$.
The number of  data for small vortices is about a third of the whole.
The ratio of $\bm v_e/\bm v_{ring}$ of the small vortex loops is distributed  between $\sim$0.88 and $\sim$1 (see the shaded part in Fig. \ref{fig:ve_v}).
The PDF of $\bm v_e/\bm v_{ring}$ of the small vortex loops has a narrower distribution than that of large ones, indicating that the small emitted vortices are closer to a circular shape than the large emitted vortices. The Kelvin waves excited on a small vortex have a large wave number and a high energy. The small vortices consequently excite  fewer Kelvin waves  than the large vortices.

\begin{figure}[h]
\begin{tabular}{cc}
\begin{minipage}{0.5\hsize}
\begin{center}
\includegraphics[clip, width=5.5cm]{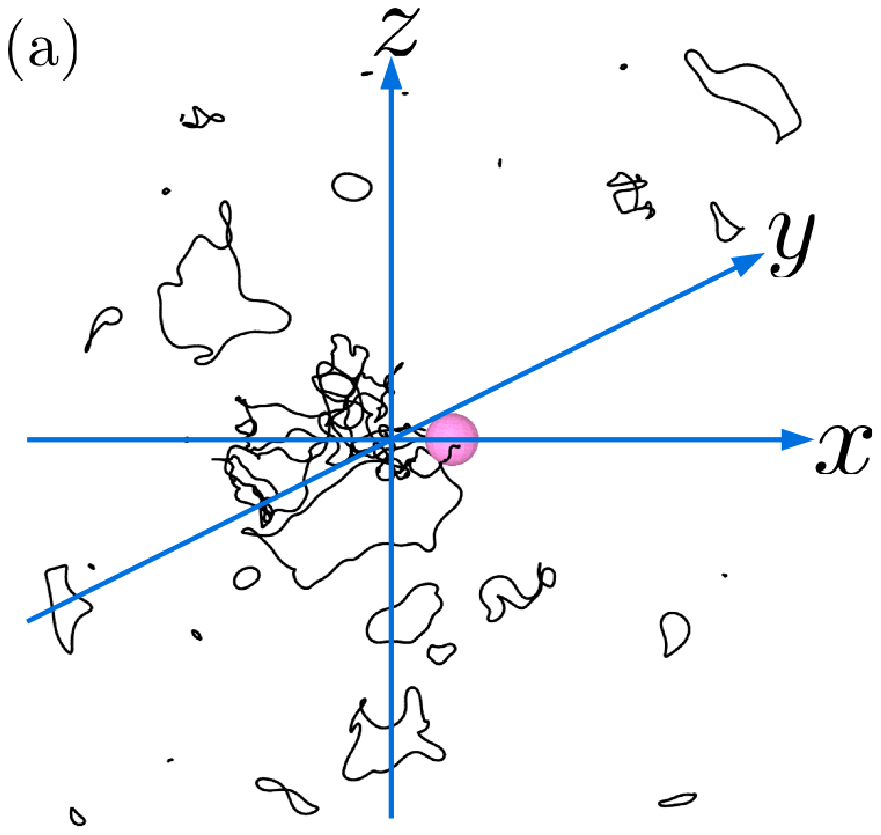}
\end{center}
\end{minipage}
\begin{minipage}{0.5\hsize}
\begin{center}
\includegraphics[clip, width=5.5cm]{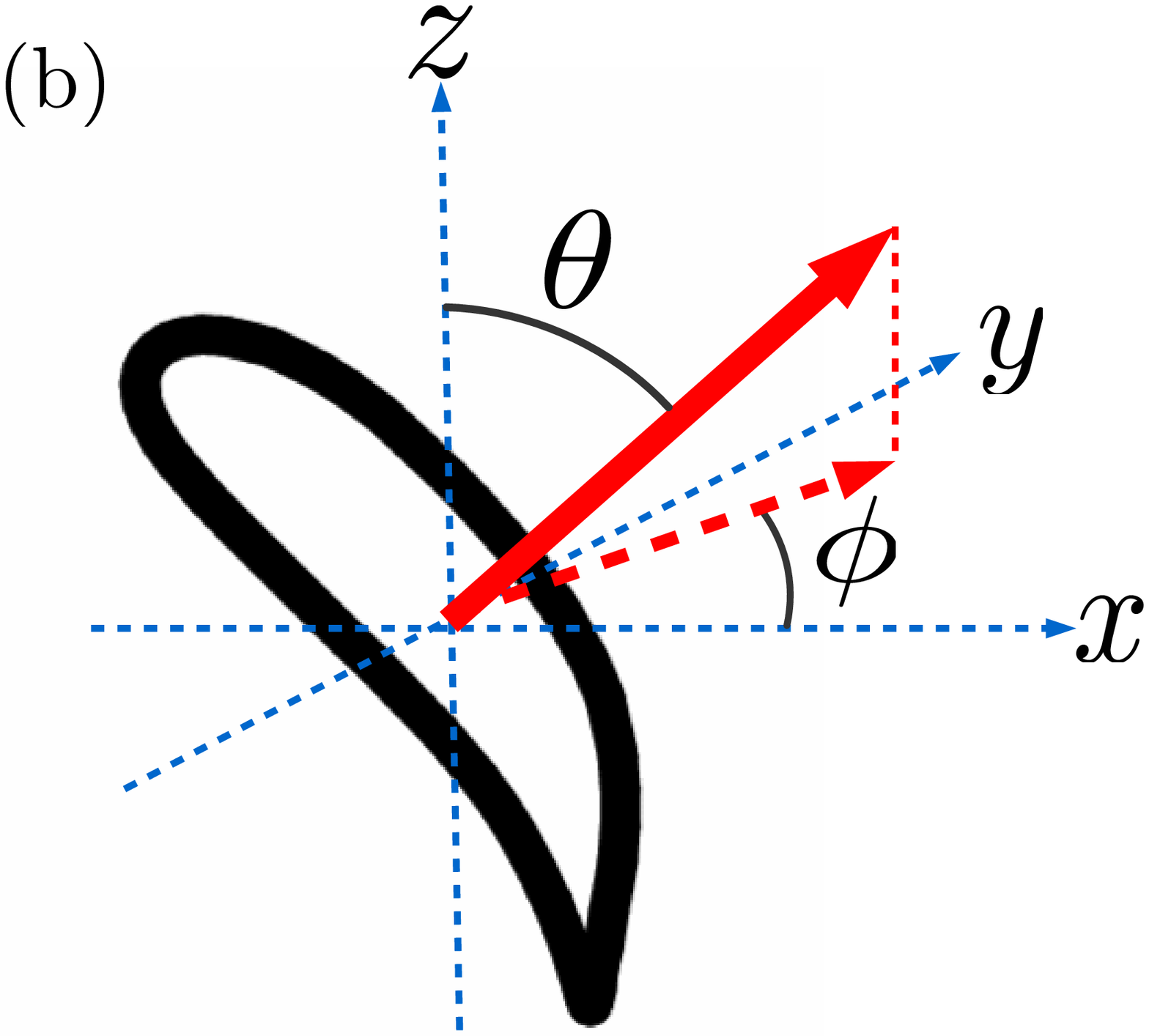}
\end{center}
\end{minipage}
\end{tabular}
\caption{(Color online) (a) The coordinate system. (b) The azimuthal angle $\phi$ of the direction of the drift velocity of a vortex loop. The red arrow shows the drift velocity vector. The dashed red arrow shows the projection of the vector on the {\it x-y} plane.}
\label{fig:origin}
\end{figure}

\begin{figure}[ht!]
\begin{center}
\includegraphics[clip, width=6.2cm]{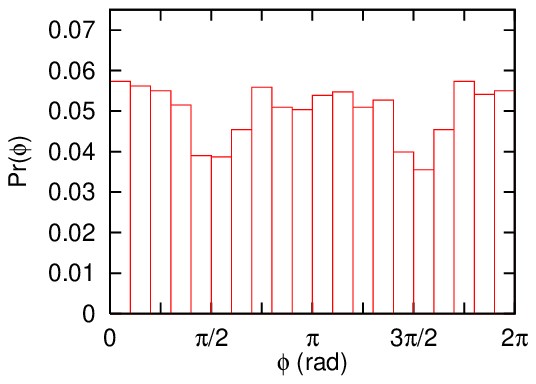}
\end{center}
\caption{(Color online) The PDF of the directions of drift velocity of emitted vortex loops with the polar angle $\pi /4 \leq \theta \leq 3\pi/4$ as a function of the azimuthal angle $\phi$.}
\label{fig:an11}
\end{figure}

\begin{figure}[ht!]
\begin{center}
\includegraphics[clip, width=6.2cm]{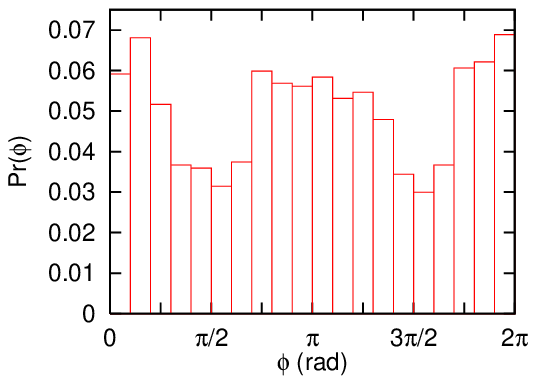}
\end{center}
\caption{(Color online) The PDF of the directions of drift velocity of emitted vortex loops whose lengths are longer than $l_{SV}$ (5.18 $\mu$m) with the polar angle $\pi /4 \leq \theta \leq 3\pi/4$ as a function of the azimuthal angle $\phi$.}
\label{fig:an11l}
\end{figure}

\begin{figure}[ht!]
\begin{center}
\includegraphics[clip, width=6.2cm]{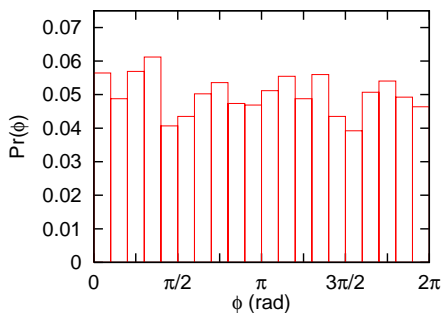}
\end{center}
\caption{(Color online) The PDF of the directions of drift velocity of emitted vortex loops whose lengths are shorter than $l_{SV}$ (5.18 $\mu$m) with the polar angle $\pi /4 \leq \theta \leq 3\pi/4$ as a function of the azimuthal angle $\phi$.}
\label{fig:an11s}
\end{figure}

\subsection{\label{sec:pdf_dir}Anisotropy of propagation direction}
This system exhibits anisotropy caused by the oscillating sphere. The propagation of the emitted vortex loops is anisotropic along the direction of the sphere oscillation in the statistically steady state.
We calculate the direction of the drift velocity, Eq. (\ref{eq:vel_all}), of each vortex loop averaged for 6 ms after the vortex reaches the boundary of the computational spherical region.

To express the propagation directions of each vortex loop, we choose the following system of coordinates:
The origin is set at the center of the oscillation, the {\it x} axis is taken along the direction of oscillation, and the {\it y} and {\it z} axes are set as shown in Fig. \ref{fig:origin}(a). The direction of the drift velocity can be represented by an azimuthal angle $\phi$ and a polar angle $\theta$ (see Fig. \ref{fig:origin}(b) ). This system is symmetric about the {\it x} axis, and we consider the range of $\pi /4 \leq \theta \leq 3\pi/4$.
We express the PDF as a function of the azimuthal angle $\phi$.

Figure \ref{fig:an11} shows the PDF of all vortices in the simulation.
The distributions of Fig. \ref{fig:an11} show that more vortex loops propagate along the oscillation direction.
Since the PDF can depend on the size of vortices, we divide the PDF into contributions from large vortex loops and from small ones.
A similar but more intense characteristic distribution is seen in Fig. \ref{fig:an11l}, which shows the PDF of large vortex loops whose length is greater than $l_{SV}$.
In contrast, as shown in Fig. \ref{fig:an11s}, the PDF of small vortex loops whose length is less than $l_{SV}$ indicates a less characteristic distribution than depicted in Fig. \ref{fig:an11}.

This change of distributions occurs at the length $l_{SV}$ related to the sphere velocity.
There are generally two mechanisms causing the anisotropy (see Sec. \ref{sec:intro}). In this simulation, we removed the mechanism attributed to the remnant vortices. Hence, the anisotropy is caused by the sphere oscillation.
Small vortex loops are generated through some reconnections. Such reconnections can occur at any time in a turbulent region independent of the sphere oscillation. As a result, the oscillating sphere has little effect on the generation of small vortex loops, with their propagation being close to isotropic.
In contrast, large vortex loops are generated through some combination type of reconnections and stretching processes. These reconnections occur in the turbulent region where the tangle is anisotropic as seen in Sec. \ref{sec:ap} and the stretching is caused by the sphere oscillation.
As a result, the oscillating sphere directly affects the propagation of large vortex loops, making it anisotropic. The same characteristic change occurs in the PDF as a function of the polar angle $\theta$.

\begin{figure}[ht!]
\begin{center}
\includegraphics[clip, width=6cm]{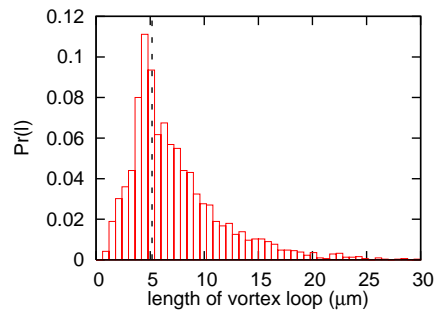}
\end{center}
\caption{(Color online) The PDF of the length of emitted vortex loops in the simulation with an oscillating sphere whose radius is 1.5 $\mu$m. The dashed black line shows the length $l_{SV}$ (5.18 $\mu$m).}
\label{fig:pdf15}
\end{figure}

\begin{figure}[ht!]
\begin{center}
\includegraphics[clip, width=6.2cm]{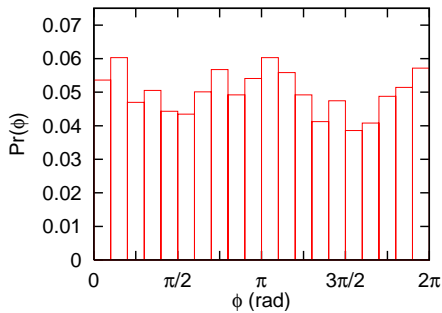}
\end{center}
\caption{(Color online) The PDF of the directions of drift velocity of emitted vortex loops with the polar angle $\pi /4 \leq \theta \leq 3\pi/4$ as a function of the azimuthal angle in the simulation with an oscillating sphere whose radius is 1.5 $\mu$m.}
\label{fig:an15}
\end{figure}

\section{\label{sec:1.5}STATISTICS FOR OTHER PARAMETERS}
In this section, we discuss the dependence of the statistics on parameters of the oscillating sphere. The parameters are the amplitude, the frequency, and the radius. Because the velocity is obtained as the product of the amplitude and the frequency, we need not consider it. The dependence on the amplitude is seen in the maximum size of the emitted vortices (see Sec. \ref{sec:pdf_len}). The dependence on the frequency affects the wavelength of the Kelvin waves excited on the vortices.
However, the effects of small changes of the frequency are negligible (see Sec. \ref{sec:vel}). Thus, we discuss the dependence on the radius of an oscillating sphere.

We also perform the same simulations with an oscillating sphere with a radius of 1.5 $\mu$m. Even if we change the radius, we could find little difference in the statistics.

The VL shows the same variation as shown in Sec. \ref{sec:vl}. The statistically steady state starts at 5.2 ms.
The average value of the anisotropic parameters in the turbulent region are $\bm I_{\parallel}=0.7076$ and $\bm I_{\bot}=0.6460$ in the statistically steady state.
We also investigate the statistics of each emitted vortex loop in the statistically steady state. Figure \ref{fig:pdf15} shows the PDF of the length of such emitted vortex loops. The PDF has a peak corresponding to the length $l_{SV}$. There is little difference in the simulations with different radii of the oscillating sphere; therefore, the distribution seems to be determined by the length $l_{SV}$ rather than the radius.
Figure \ref{fig:an15} shows the PDF of the propagation direction as a function of the azimuthal angle $\phi$,  which exhibits the same characteristic distribution as in Fig. \ref{fig:an11}. The PDF also depends on the length of the emitted vortex loops. The same characteristic distribution is seen in the PDF of large vortices whose length is greater than the length $l_{SV}$ in the simulation. In contrast, the PDF of small vortex loops whose length is shorter than the length $l_{SV}$ has no characteristic distribution.

In this simulation, changing the amplitude of the oscillation can affect the length of emitted vortex loops.
Based on our study, the statistics in this simulation are insensitive to variation in the other parameters such as the sphere radius and the frequency.
\\

\section{\label{sec:conclusion}CONCLUSIONS}
We numerically studied quantum turbulence created by an oscillating sphere. Seed vortices were continuously injected onto the surface of the sphere and stretched by the sphere motion to form a localized vortex tangle, which emits vortex loops. We investigated the statistics of the emitted loops, and by considering QT to consist of quantized vortices, we investigated the statistics of each vortex. The supposed situation is realized in the OCU experiments using vibrating wires in which a generator wire created a vortex tangle from remnant vortices and the vortices emitted from the tangle are monitored by a detector wire.

The emitted vortex loops show some characteristic properties, though some influence of the injected vortices remains.
The distribution of the length of vortex loops has a peak at the length related to the sphere velocity and covers a wide range of values because of reconnection and stretching. The drift velocity of the emitted vortex loop is lower than that of a circular vortex ring of the same length. The propagation direction of the emitted vortex loops is anisotropic along the oscillation direction of the sphere; more vortices propagate along the oscillation direction than along the direction normal to it.
Simulations for sphere radii of 1.1 and 1.5 $\mu$m reveal little difference in the statistics. Hence, the effective parameters responsible for the statistics can be the amplitude and the velocity of the sphere oscillation.

Our final goal would be to ascertain information on  the localized vortex tangle from the statistics of the emitted vortices. Of course, it is impossible to reproduce the thorough information of the tangle, but some information can be surmised. The anisotropic propagation of the emitted vortices suggests that the configuration of the tangle around the sphere is anisotropic as well. The anisotropic parameters actually demonstrate that the vortex tangle around the sphere is slightly anisotropic along the oscillation direction. The statistics of the emitted vortex length should come from those of vortices inside the localized tangle, but we do not know their proper correspondence.

In experiments of quantum turbulence using vibrating objects, it will be worth investigating such statistical laws of each vortex described in this article.

\begin{acknowledgments}
We  thank T. Hata and Y. Nago for helpful discussions. This research was supported by a Grant-in-Aid for Scientific Research (B) (Grant No. 23340108) from the Japan Society for the Promotion of Science.
\end{acknowledgments}


\begin{thebibliography}{99}
\bibitem{book}
W. P. Halperin and M. Tsubota, {\it Progress in Low Temperature Physics} (North-Holland, Amsterdam, 2009), Vol.16.
\bibitem{remnant1}
N. Hashimoto, R. Goto, H. Yano, K. Obara, O. Ishikawa, and T. Hata,  Phys. Rev. B \textbf{76}, 020504 (2007).
\bibitem{Kubo}
H. Kubo, H. Yano, Y. Nago, A. Nishijima, K. Obara, O. Ishikawa, and T. Hata, J. Low Temp. Phys. \textbf{171}, 466 (2012).
\bibitem{Yano}
Y. Nago, A. Nishijima, H. Kubo, T. Ogawa, K. Obara, H. Yano, O. Ishikawa, and T. Hata,  Phys. Rev. B \textbf{87}, 024511 (2013).
\bibitem{bigsph}
R. H$\ddot{\rm a}$nninen, M. Tsubota, and W. F. Vinen, Phys. Rev. B \textbf{75}, 064502 (2007).
\bibitem{Fuji2008}
R. Goto, S. Fujiyama, H. Yano, Y. Nago, N. Hashimoto, K. Obara, O. Ishikawa, M. Tsubota, and T. Hata, Phys. Rev. Lett. \textbf{100}, 045301 (2008).
\bibitem{Fuji2009}
S. Fujiyama and M. Tsubota, Phys. Rev. B \textbf{79}, 094513 (2009).
\bibitem{sch85}
K. W. Schwarz, Phys. Rev. B \textbf{31}, 5782 (1985).
\bibitem{sch88}
K. W. Schwarz, Phys. Rev. B \textbf{38}, 2398 (1988).
\bibitem{image}
P. G. Saffman, {\it Vortex Dynamics} (Cambridge University Press, Cambridge, England,  1992)
\bibitem{sch74}
K. W. Schwarz, Phys. Rev. A \textbf{10}, 2306 (1974).
\bibitem{araki}
M. Tsubota, T. Araki, and S. K. Nemirovskii, Phys. Rev. B \textbf{62}, 11751 (2000).
\bibitem{donelly}
R. J. Donnelly, {\it Quantized Vortices in Helium II} (Cambridge University Press, Cambridge, England, 1991).
\bibitem{PRE}
C. F. Barenghi, R. H$\ddot{\rm a}$nninen, and M. Tsubota, Phys. Rev. E \textbf{74}, 046303 (2006).
\bibitem{kelvin}
E. B. Sonin, Europhys. Lett. \textbf{97}, 46002 (2012).
\end{thebibliography}
\end{document}